\documentclass[conference]{IEEEtran}
\IEEEoverridecommandlockouts

\usepackage{mathtools}
\usepackage[T1]{fontenc}
\usepackage{textcomp}
\usepackage{algpseudocode}
\usepackage{algorithm}
\usepackage{xcolor}
\usepackage{bytefield}
\usepackage{lipsum}
\usepackage{float}
\usepackage{setspace}
\usepackage{url}
\usepackage{caption}
\usepackage[pdftex]{graphicx}
\usepackage{multirow}
\usepackage{soul}
\usepackage{balance}

\usepackage{array}
\usepackage[bottom]{footmisc}
\newcolumntype{P}[1]{>{\centering\arraybackslash}p{#1}}

\begin{document}
\title{Reducing Data Motion to Accelerate the Training of Deep Neural Networks} 
\author{\IEEEauthorblockN{Sicong Zhuang}
\IEEEauthorblockA{\textit{Computer Sciences Department} \\
\textit{Barcelona Supercomputing Center (BSC)}\\
Barcelona, Spain \\
sicong.zhuang@bsc.es}
\and
\IEEEauthorblockN{Cristiano Malossi}
\IEEEauthorblockA{\textit{Foundations of Cognitive Solutions} \\
\textit{IBM Zurich Research Laboratory}\\
Zurich, Switzerland \\
ACM@zurich.ibm.com}
\and
\IEEEauthorblockN{Marc Casas}
\IEEEauthorblockA{\textit{Computer Sciences Department} \\
\textit{Barcelona Supercomputing Center (BSC)}\\
Barcelona, Spain \\
marc.casas@bsc.es}
}

\maketitle

\begin{abstract}
The use of Deep Neural Networks (DNNs) is becoming ubiquitous in many areas due to their exceptional pattern detection capabilities.
For example, deep learning solutions are coupled with
large scale scientific simulations to increase the accuracy of pattern 
classification problems and thus improve the quality of scientific computational models.
Despite this success, deep learning methods still incur several important 
limitations: the DNN topology must be set by going through an empirical and 
time-consuming process, the training phase is very costly and the latency of the 
inference phase is a serious limitation in emerging areas like autonomous driving.

This paper reduces the cost of DNNs training by decreasing the amount of data 
movement across heterogeneous architectures composed of several GPUs and 
multicore CPU devices.
In particular, this paper proposes an algorithm to dynamically adapt the data representation format of network weights during  
training.
This algorithm drives a compression procedure that reduces data size before sending them over the parallel system.
We run an extensive evaluation campaign considering several up-to-date deep neural network 
models and two high-end parallel architectures composed of multiple GPUs and CPU multicore chips.
Our solution achieves average performance improvements from 6.18\% up to 11.91\%.
\end{abstract}


\begin{IEEEkeywords}
Approximate Computing, Heterogeneous Parallel Systems, Deep Learning, Convolutional Neural Networks
\end{IEEEkeywords}

\section{Introduction}

The use of Deep Neural Networks~(DNNs) is becoming ubiquitous in areas like 
computer vision (e.g., image recognition and object detection)~\cite{alexnet, 
Inception}, speech recognition~\cite{Acoustic}, language 
translation~\cite{Language}, and many more~\cite{Ciregan2012}.  DNNs provide 
very competitive pattern detection capabilities and, more specifically, 
Convolutional Neural Networks~(CNNs) classify very large image sets with 
remarkable accuracy~\cite{Krizhevsky2012}.
Indeed, DNNs already play a very significant role in the large production 
systems of major IT companies and research centers, which has in turn driven the 
development of advanced software frameworks for the deep learning 
area~\cite{tensorflow} as well as DNN-specific hardware 
accelerators~\cite{Merolla668,Jouppi2017}.  As an example, deep learning 
solutions are being coupled with physical computational models  for solving 
pattern classification problems in the context of large-scale climate 
simulations~\cite{Kurth2017}.
Despite all these accomplishments, deep learning models still suffer from 
several fundamental problems: the neural network topology is determined through 
a long and iterative empirical process, the training procedure has a huge cost 
in terms of time and computational resources, and the inference process of large 
network models incurs considerable latency to produce an output, which is not 
acceptable in domains requiring real-time responses like autonomous driving.

To deal with the large quantity of Floating Point computations required to train a 
DNN, GPUs are usually employed~\cite{You17}.  They exploit the large amount of 
data-level parallelism of deep learning workloads.  Although GPUs and other 
hardware accelerators have been successfully employed to boost the training 
process, data exchanges involving different accelerators may incur significant 
performance penalties. 

This paper describes and evaluates a method that exploits DNNs tolerance to 
data representation formats smaller than the commonly used 32-bit Floating Point 
(FP) standard~\cite{gupta15, flexpoint17}.
Our method accelerates the training of 
DNNs by reducing the cost of data transfers across heterogeneous high-end 
architectures integrating multiple GPUs without deterioration on the training 
accuracy.
Our solution is designed to efficiently use the incoming bandwidth of the GPU 
accelerators. It relies on an adaptive scheme that dynamically adapts the data 
representation format required by each DNN layer and compresses network 
parameters before sending them over the parallel system. 
This paper makes the following contributions:

\begin{itemize}
\item It proposes the {\it Adaptive Weight Precision (AWP)} algorithm, which 
    dynamically adapts the numerical representation of DNN weights during 
        training. AWP relies on DNNs' tolerance for reduced data representation 
        formats. It defines the appropriated data representation format per 
        each network layer during  training without hurting network accuracy.

\item It proposes the {\it Approximate Data Transfer (ADT)} procedure to 
        compress DNN's weights according to the decisions made by the AWP algorithm.  
        ADT relies on both thread- and SIMD-level parallelism  and is compatible 
        with architectures like IBM's POWER or x86. ADT is able to compress 
        large sets of weights with minimal overhead, which enables the large 
        performance benefits of our approach.

\item It evaluates ADT and AWP on two high-end systems: The first is composed of 
        two x86 Haswell multicore devices plus four NVIDIA Tesla GK210 GPU 
        accelerators and the second system integrates two POWER9 chips and four 
        NVIDIA Volta V100 GPUs.  Our evaluation considers the 
        Alexnet~\cite{alexnet}, the VGG~\cite{vgg} and the Resnet~\cite{resnet} 
        network models applied to the ImageNet ILSVRC-2012 
        dataset~\cite{imagenet}. Our experiments report average performance 
        benefits of 6.18\% and 11.91\% on the x86 and the POWER systems, 
        respectively.
\end{itemize}

Many proposals describe how data representation formats smaller than the 32-bit 
Floating Point IEEE standard can be applied to deep learning workloads without 
hampering their accuracy~\cite{bottou08, gupta15, Micikevicius2018}.
This paper presents the first approach that uses reduced data formats to 
minimize data movement during DNN training in heterogeneous high-end systems,
which are extensively used to run deep learning workloads~\cite{You17}.

This paper is structured as follows:
Section~\ref{sec:adaptive} describes our first contribution, the Adaptive Weight 
Precision algorithm (AWP).  Section~\ref{sec:approx} details the Approximate 
Data Transfer (ADT) procedure.  Section~\ref{sec:setup} explains the 
experimental setup of this paper.  Section~\ref{sec:evalutation} describes the  
experiments we conduct to evaluate AWP and ADT on three state-of-the-art neural 
networks.  Section~\ref{sec:Relatedworks} describes the most relevant related 
work.  Finally, Section~\ref{sec:conclusion} summarizes the conclusions of 
this paper.




\section{The Adaptive Weight Precision (AWP) Algorithm}
\label{sec:adaptive}

The Adaptive Weight Precision (AWP) algorithm relies on the tolerance of DNNs to data representation formats smaller than the 32-bit Floating Point standard.
Indeed, previous work indicates that,
unlike scientific codes focused on solving partial differential equations or large linear systems,
neural networks do not always require 32-bit representation during training~\cite{bottou08, gupta15}. 
Even more, adding stochastic noise to certain variables during the learning phase
improves DNNs accuracy~\cite{murray94, bishop95, aud13}.
Nevertheless, when facing unknown scenarios in terms of new workloads or parameter settings,
the data representation requirements of DNNs 
are non-trivial to be determined and, to make things more complicated, they may change as the training phase progresses.

The AWP algorithm dynamically  
determines data representation requirements per each network layer by monitoring the 
evolution of the $l^2$-norm of the weights.
AWP identifies the number of bits required to represent 
DNNs weights and guarantees the progress of the training process.
AWP assigns the same data representation format to all weights 
belonging to a certain network layer.
The training starts with a small data representation that is independently 
increased for each layer. 

\begin{algorithm}
\caption{Adaptive Weight Precision (AWP) Algorithm}
\label{alg:norm}
{\fontsize{8}{8}\selectfont
\begin{algorithmic}[1]
    \State BitsPerLayer := [$B_0, B_1, \hdots, B_{NumLayers}$]
    \Comment List storing the number of bits corresponding to the data representation of each layer
    \State IntervalCounter := [0, 0, $\hdots$, 0]
    \Comment List storing the number of times the relative change rate
             fails to meet the threshold per layer
    \For {batch := 0 \ldots NumBatches}
        \State Apply backpropagation to batch
        \For {layer := 0 \ldots NumLayers}
            \State $\delta := \frac{(|\text{W}_{batch, layer}| - |\text{W}_{batch-1, layer}|)}{|\text{W}_{batch-1, layer}|}$
            \If {$\delta$ < T}
                \State $\text{IntervalCounter}_{layer}$ +=  1
            \EndIf
            \If {$\text{IntervalCounter}_{layer}$ == INTERVAL}
                \State $\text{BitsPerLayer}_{layer}$ += N
                \State $\text{IntervalCounter}_{layer}$ := 0
            \EndIf
        \EndFor
        \EndFor
\end{algorithmic}
}
\end{algorithm}

Algorithm~\ref{alg:norm} displays a pseudo-code description of AWP.
Once the backpropagation process has been applied to a given batch, AWP iterates over all network layers.
The algorithm computes per each batch and network layer the $l^2$-norm of all its weights' values 
and derives the relative change rate~$\delta$ of the $l^2$-norm 
with regard to the previously processed batch. 
For the batch $i$, the change rate is defined as  
$\delta_i=(|W_i| - |W_{i-1}|)/|W_{i-1}|$, where $W_i$ is the vector of weights 
of a certain layer while batch $i$ is processed. 
Every time the change rate is below a given threshold \textit{T} for a certain 
layer, the algorithm accounts for it by increasing the $IntervalCounter$ parameter.
The algorithm increases $N$~bits of precision if the change 
rate is below \textit{T} during a certain number of batches defined by the parameter
\textit{INTERVAL} 
and sets the $IntervalCounter$ parameter of the corresponding layer to zero.
Section~\ref{sec:evaluation1} describes how we determine the values of parameters \textit{T}, \textit{INTERVAL}, and \textit{N}.

\section{The Approximate Data Transfer (ADT) Procedure}
\label{sec:approx}

The Approximate Data Transfer (ADT) procedure compresses network's weights
before they are transferred to the GPUs.
In the context of DNNs training on heterogeneous multi-GPU nodes, CPU multicore devices 
are typically responsible for orchestrating the parallel run and updating DNN parameters.
Once the process of a batch starts, the updated parameters including the weights $W$ are sent to each GPU.
If the set of parameters does not fit in GPUs' main memory, they are sent on several phases as the different GPUs need them.
The different samples of each batch are evenly distributed across all GPUs.
Therefore, each GPU computes its contribution to the gradient $\Delta W$ by processing its corresponding set of samples.
The CPU multicore subsequently gathers all contributions to the gradient and 
uses them for weight updates  
$W \gets W - \mu (\frac{1}{n} \sum_{i}^{n} \Delta W_{i})$, where $\mu$ is 
the learning rate.

Data movement involving different GPU devices increases as the network topology becomes more complex or the number of training samples grows, which can saturate the system bandwidth and become a major performance bottleneck.
This paper mitigates this issue by compressing network weights before they are sent to the GPU devices.
The AWP algorithm described in Section~\ref{sec:adaptive} determines, for all  
weights belonging to a particular network layer, the number of bits to send.
In this context, to efficiently compress and decompress network weights, ADT uses of two procedures that constitute its fundamental building blocks. 
These procedures are complementary and applied either before or after data transfers to GPUs.
\begin{itemize}
    \item \textbf{Bitpack} compresses the weights discarding the less significant bits on the CPU side;
    \item \textbf{Bitunpack} converts the weights back to the IEEE-754 32-bit Floating Point format on the GPUs. 
\end{itemize}

\begin{figure}
    \centerline{\includegraphics[scale=0.30]{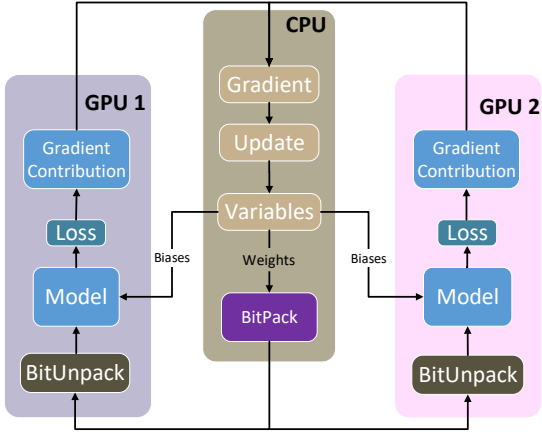}}
        \caption{The ADt on a 2-GPU system. Variables include: weights which go through
    the ADt procedure and biases which are sent directly to the GPUs to build the
    network model together with the unpacked weights.}
        \label{bitpack_mgpu}
\end{figure}

Figure~\ref{bitpack_mgpu} provides an example including a multicore CPU and two GPU devices to describe the way both Bitpack and Bitunpack procedures operate.
All neural network parameters (weights and biases) are updated at the CPU level, which is where the Bitpack procedure takes place. 
We do not apply the Bitpack procedure to the network biases since we do not observe any significant performance benefit from compressing them.
Since each output neuron requires just one bias parameter, the total number of them is significantly smaller than the total number of weights.
At the beginning of each SGD iteration the compressed weights are sent to each GPU together with the biases and the corresponding training samples.
Each GPU uncompresses the weights, builds the neural network model,  
and computes its specific contribution to the gradient.
These contributions are sent to the CPU, which gathers them, computes the gradient, and updates network parameters.

The Bitpack operation runs on CPU multicore devices.
To boost Bitpack  
we use OpenMP~\cite{openmp} and 
Single-Instruction Multiple Data (SIMD) intrinsics.
OpenMP is used to run Bitpack on several threads.
The use of SIMD instructions allows Bitpack to operate at the SIMD register level, which
avoids incurring large performance penalties in the process of producing the reduced-size weights.
We implement two versions of Bitpack.
One version uses Intel's AVX2~\cite{avx} instruction set and the other one relies on AltiVec~\cite{Altivec}. 
Bitpack can be implemented on top of any SIMD instruction set architecture supporting simple byte shuffling instructions at the register level.

The Bitunpack procedure runs on the GPUs.
It can be trivially parallelized since each weight is mapped
to a single 32-bit FP variable, which means that GPUs can process a
large amount of weights simultaneously and efficiently build the DNN model.
In fact, Bitunpack incurs negligible overhead as Section~\ref{sec:performance} shows.

ADT manipulates the internal representation of network weights by discarding some bits.
We use the standard 32-bit IEEE-754 single-precision Floating Point format
~\cite{ieee754} (1 bit sign, 8 bits exponent and 23 bits mantissa) for all the computation routines.
The Bitpack method considers network weights as 32-bit 
words where rounding to $N$ bits means discarding the lowest $32-N$ bits.

\begin{algorithm}
\caption{High Level Pseudo-code Version of Bitpack}
\label{alg:bitpack_na\"{i}ve}
{\fontsize{9}{9}\selectfont
\begin{algorithmic}[1]
    \State W
    \Comment Array of 32-bit Floating Point values containing weights
    \State Pw
    \Comment Array containing the reduced precision weights
    \State RoundTo
    \Comment Number of bytes to keep per weight
    \State POffset := 0
    \Comment Indicates the current size (in bytes) of Pw
    \For {weight \textbf{in} W}
        \State Pw[POffset : POffset+RoundTo] := weight[0 : RoundTo]
        \Comment Copy most significant RoundTo bytes to Pw
        \State POffset := POffset + RoundTo
    \EndFor
\end{algorithmic}
}
\end{algorithm}

\subsection{Bitpack}
\label{subsec:bitpack}
A high-level version of the Bitpack procedure in terms of pseudo-code is illustrated by 
Algorithm~\ref{alg:bitpack_na\"{i}ve}.
The algorithm requires a couple of arrays: the input array $W$, which contains all the weights of a certain network layer, and an 
output array $Pw$, which stores the compressed versions of these weights. 
The algorithm goes through 
the entire $W$ input array and, per each weight, it copies the most 
significant $RoundTo$ bytes to the output array $Pw$.
Our Bitpack implementation manipulates data at the byte granularity.
We do not observe significant performance benefits when operating at finer granularity in the experiments we run.
The AWP algorithm described in Section~\ref{sec:adaptive} determines the data representation format per each network layer.  
The number of bits of the chosen format is rounded to the nearest number of bytes
that retains all of its information (E.g., if AWP provides the value 14, $RoundTo$ will be set to 2 bytes).
The $Pw$ array is sent to the GPUs once the Bitpack procedure finishes compressing network weights.

Deep networks usually contain tens or even hundreds of millions of 
weights~\cite{alexnet, alexnet2, vgg}, which makes any trivial implementation 
of Algorithm~\ref{alg:bitpack_na\"{i}ve} not applicable in practice.
We mitigate compression costs by observing that Algorithm~\ref{alg:bitpack_na\"{i}ve} is trivially parallel since processing one weight just requires the $RoundTo$ parameter.
Algorithm~\ref{alg:bitpack_omp} shows how to parallelize the Bitpack procedure by using OpenMP threads.
Each thread takes care of a certain portion of the $Pw$ array.

\begin{algorithm}
\caption{Bitpack with OpenMP}
\label{alg:bitpack_omp}
{\fontsize{9}{9}\selectfont
\begin{algorithmic}[1]
    \State W 
    \Comment Array of 32-bit Floating Point values containing weights
    \State Pw
    \Comment Array containing the reduced precision weights
    \State RoundTo 
    \Comment Number of bytes to keep per weight 
    \State NumThreads
    \Comment Number of OpenMP threads
    \State \textcolor{orange} {\#pragma omp parallel for} 
      \For {weight \textbf{in} W}
            \State POffset := Corresponding position in Pw
            \State Pw[POffset : POffset+RoundTo] := weight[0 : RoundTo]
            \Comment Copy the most significant RoundTo bytes to Pw
      \EndFor
\end{algorithmic}
}
\end{algorithm}

\subsection{Single Instruction Multiple Data Bitpack}
Since all weights within one layer are processed in the same way by the Bitpack procedure, we can leverage Single Instruction Multiple Data (SIMD) instructions to vectorize it.
Most state-of-the-art architectures implement SIMD instruction set: IBM's 
AltiVec~\cite{Altivec}, Intel's Advanced Vector Extensions (AVX)~\cite{avx}, and ARM's Neon~\cite{neon}.
In our experiments we use
Intel's AVX2~\cite{avx}, which implements a set of SIMD 
instructions operating over 256-bit registers, and IBM's AltiVec instruction 
set~\cite{Altivec}, which has SIMD instructions operating over 128-bit registers.
Section~\ref{sec:setup} describes the specific details of our evaluation considering both x86 and POWER architectures.

Figure~\ref{bitpack_avx2} shows the byte-level operations of SIMD-based Bitpack applied to eight 32-bit weights and implemented with AVX2. 
The \textit{RoundTo} parameter is set to 3, which implies discarding the last 8 bits of 
each weight since the target data representation is 24-bit long. 
First, eight 32-bit Floating Point weights are loaded to a 256-bit register.
In the next step, we use \textit{\_mm256\_shuffle\_epi8} to shuffle the least significant
eight bits of each weight to the least significant bits of their respective 128-bit lane 
(see the grey area of Figure~\ref{bitpack_avx2} Step 2) and pack the rest of the bits 
together. 
Afterwards we use \textit{\_mm256\_permutevar8x32\_epi32} to do the same operation 
across the two 128-bit lanes. 
Finally, we use \textit{\_mm256\_maskstore\_epi32} to just store  
the resulting 192 bits to the target array.
Not all AVX2 instructions operate over the entire 256-bit register.
Instead, many of
them conceive the register as two 128-bit lanes and operate on them separately.
This is the reason way we can not carry out Steps 2 and 3 by using a single AVX2 instruction.

\definecolor{lightgray}{gray}{0.5}
\begin{figure}
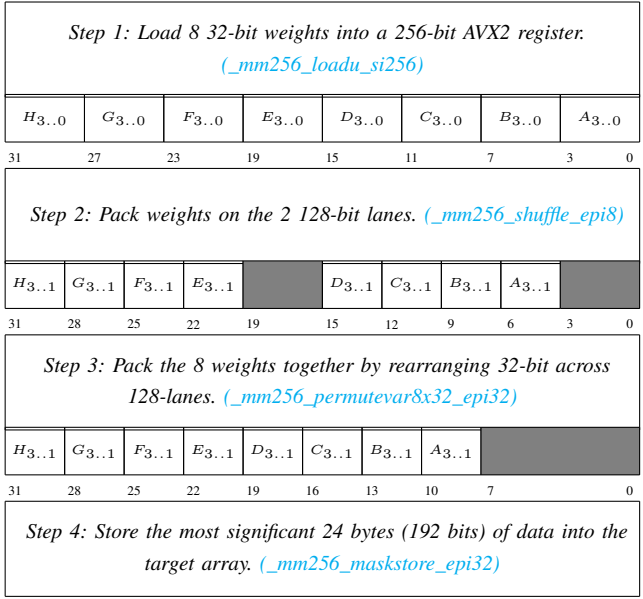

    \centering
    \begin{bytefield}[bitwidth=0.75em, bitformatting={\tiny},endianness=big]{32}
        \wordbox{2}{\textit{\fontsize{8}{8} \selectfont 
            Step 1: Load 8 32-bit weights into a 256-bit AVX2 register. 
            \textcolor{cyan}{(\_mm256\_loadu\_si256)}}} \\ 
        \bitboxes{4}{{\tiny $H_{3..0}$}&{\tiny $G_{3..0}$}&{\tiny $F_{3..0}$}&
                    {\tiny $E_{3..0}$}&{\tiny $D_{3..0}$}&
                    {\tiny $C_{3..0}$}&{\tiny $B_{3..0}$}&{\tiny $A_{3..0}$}} \\
        \bitheader{0,3,7,11,15,19,23,27,31} \\
        \wordbox{2}{\textit{\fontsize{8}{8} \selectfont
            Step 2: Pack weights on the 2 128-bit lanes.
            \textcolor{cyan}{(\_mm256\_shuffle\_epi8)}}} \\
        \bitboxes{3}{{\tiny $H_{3..1}$}} & 
        \bitboxes{3}{{\tiny $G_{3..1}$}} & 
        \bitboxes{3}{{\tiny $F_{3..1}$}} & 
        \bitboxes{3}{{\tiny $E_{3..1}$}} & 
        \bitbox{4}{\color{lightgray}\rule{\width}{\height}} &
        \bitboxes{3}{{\tiny $D_{3..1}$}} &
        \bitboxes{3}{{\tiny $C_{3..1}$}} &
        \bitboxes{3}{{\tiny $B_{3..1}$}} &
        \bitboxes{3}{{\tiny $A_{3..1}$}} &
        \bitbox{4}{\color{lightgray}\rule{\width}{\height}} \\
        \bitheader{0,3,6,9,12,15,19,22,25,28,31} \\
        \wordbox{2}{\textit{\fontsize{8}{8} \selectfont
            Step 3: Pack the 8 weights together by rearranging 32-bit across 
            128-lanes.
            \textcolor{cyan}{(\_mm256\_permutevar8x32\_epi32)}}} \\
        \bitboxes{3}{{\tiny $H_{3..1}$}} & 
        \bitboxes{3}{{\tiny $G_{3..1}$}} & 
        \bitboxes{3}{{\tiny $F_{3..1}$}} & 
        \bitboxes{3}{{\tiny $E_{3..1}$}} & 
        \bitboxes{3}{{\tiny $D_{3..1}$}} &
        \bitboxes{3}{{\tiny $C_{3..1}$}} &
        \bitboxes{3}{{\tiny $B_{3..1}$}} &
        \bitboxes{3}{{\tiny $A_{3..1}$}} &
        \bitbox{8}{\color{lightgray}\rule{\width}{\height}} \\
        \bitheader{0, 7, 10, 13, 16, 19, 22, 25, 28, 31} \\
        \wordbox{2}{\textit{\fontsize{8}{8} \selectfont
            Step 4: Store the most significant 24 bytes (192 bits) of data into 
            the target array. \textcolor{cyan}{(\_mm256\_maskstore\_epi32)}}} \\
    \end{bytefield}
	\vspace{-0.5cm}
	\caption{Bitpack implemented with AVX2, RoundTo=3}
	\label{bitpack_avx2}
\end{figure}


\begin{algorithm}
\caption{Bitpack with OpenMP + AVX2}
\label{alg:bitpack_omp_avx}
{\fontsize{9}{9}\selectfont
\begin{algorithmic}[1]
    \State W
    \Comment Array of 32-bit Floating Point values containing weights
    \State Pw
    \Comment Array containing the reduced precision weights
    \State RoundTo
    \Comment Number of bytes to keep per weight
    \State \textcolor{orange} {\#pragma omp parallel for} 
    \For {weights \textbf{in} W}
        \State \textcolor{cyan}{\_mm256\_loadu\_si256}
        \Comment Load 8 32-bit weights 
        \State \textcolor{cyan}{\_mm256\_shuffle\_epi8}
        \Comment Compress at each 128-bit lane
        \State \textcolor{cyan}{\_mm256\_permutevar8x32\_epi32}
        \Comment Shuffle the compressed weights into the most significant bits
        \State \textcolor{cyan}{\_mm256\_maskstore\_epi32}
        \Comment Store compressed weights to the target array
    \EndFor
\end{algorithmic}
}
\end{algorithm}

\begin{algorithm}
\caption{Bitunpack on GPU}
\label{alg:bitunpack}
{\fontsize{9}{9}\selectfont
\begin{algorithmic}[1]
    \State Pw
    \Comment Array containing compressed weights
    \State W
    \Comment Array of 32-bit Floating Point values containing weights
    \State RoundTo
    \Comment The number of bytes that are going to be kept
    \For{UnitId := 0 \ldots NumUnit}
    \State Distribute W and Pw across all the computation units in the GPU
        \State POffset := 0
        \For{weight in W}
            \State weight := Pw[POffset : POffset+RoundTo] $\ll$ (4 - RoundTo) * 8
            \State POffset := POffset + RoundTo
        \EndFor
    \EndFor
\end{algorithmic}
}
\end{algorithm}

Algorithm~\ref{alg:bitpack_omp_avx} summarizes our implementation of the Bitpack procedure with AVX2.
It exploits two-level parallelism: first, the input array of weights is distributed across several threads.
Second, within each thread, the compression of each eight 32-bit weights subset is performed at the register level by means of byte shuffling instructions.
This sophisticated procedure exploiting parallelism at both thread and SIMD register levels uses all the available hardware resources and avoids costly memory accesses.

\subsection{Bitunpack}
Once data in reduced-size format reaches the target GPU, the Bitunpack procedure immediately 
restores them into their original IEEE-754 32-bit Floating Point format. 
We display pseudo-code describing this process in Algorithm~\ref{alg:bitunpack}.
Bitunpack reads the reduced-sized weights from array $Pw$ and assigns additional bits to them. 
Bitunpack gives zero values to these additional bits.
We distribute the Bitunpack process across the whole GPU, which enables an extremely parallel scheme exploiting GPUs manycore architecture. 

The Bitunpack routine is developed using CUDA~\cite{cuda}. 
Our code  
runs in parallel on $N$ CUDA threads and the CUDA runtime handles the dynamic mapping between threads and the underlying GPU compute units.
Since each thread involved in the parallel run targets a different portion of the $Pw$ array, our Bitunpack procedure exposes a large amount of parallelism to the numerous computing units integrated into high-end GPU devices.

\section{Experimental Setup}
\label{sec:setup}

The experimental setup considers a large image dataset, three state-of-the-art 
neural network models and two high-end platforms.
The following sections describe all these elements in detail.

\subsection{Image Dataset}
We consider the ImageNet ILSVRC-2012 dataset~\cite{imagenet}.
The original ImageNet dataset includes three sets of images of 1000 classes 
each:
training set (1.3 million images), validation set (50,000 images) and
testing set (100,000 images).
We consider a subset of 200 classes for the wide evaluation we show in Sections~\ref{sec:alexnet},~\ref{sec:VGG},~\ref{sec:Resnet}, and~\ref{sec:Average}, which considers three different network models, three different batch sizes per model, two different platforms and three different training approaches.
Considering 1000 classes makes the training process around 170 hours long, which 
is prohibitively expensive for this large experimental campaign.  
We consider the whole ImageNet data set in the experiments we show in Section~\ref{sec:ImageNet1000}, which confirm the trends observed when considering the reduced data set.
For the rest of this paper, we refer to the 200 and 1000 classes datasets as 
ImageNet200 and ImageNet1000, respectively.
Since it is a common practice~\cite{vgg}, we evaluate the ability of a certain 
network in properly dealing with the ImageNet ILSVRC-2012 dataset in terms of 
the top-5 validation error computed over the validation set.

\subsection{DNN Models and Training Parameters}
\label{sec:trainingparameters}
We apply the AWP algorithm along with the ADT procedure on three 
state-of-the-art DNN models: a modified version of Alexnet~\cite{alexnet} with 
an extra fully-connected layer of size 4096, the configuration A of the VGG 
model~\cite{vgg} and the Resnet network~\cite{resnet}.  All hidden layers are 
equipped with a Rectified Linear Units (ReLU)~\cite{alexnet}.
The exact configurations of the three neural networks are shown in 
Table~\ref{table:config}.  The Alexnet model is composed of 5 convolutional 
layers and 4 fully-connected ones, VGG contains 8 convolutional layers and 3 
fully-connected ones and Resnet is composed of 33 convolutional layers and a 
single fully-connected one.

\begin{table}[]
\caption{Neural network configurations: The convolutional layer parameters are 
    denoted as ``conv<receptive field size>-<number of channels>''. The ReLU 
    activation function is not shown for brevity. The building blocks of Resnet
    and the number of times they are applied are shown in a single cell.}
    \centering
    \begin{tabular}{|P{2.5cm}|P{2.5cm}|P{2.5cm}|}
    \hline
    \textbf{Alexnet} & \textbf{VGG}  & \textbf{Resnet-34}  \\ \hline
    \multicolumn{3}{|c|}{input(224x224 RGB image)} \\ \hline
    conv11-64  & conv3-64 & conv7-64 \\ 
    \hline
    \multicolumn{3}{|c|}{maxpool} \\ \hline
    conv5-192 & conv3-128 & 
        \begin{tabular}[c]{@{}c@{}}conv3-64\\ conv3-64\\ x3\end{tabular}   \\ 
            \hline
\multicolumn{2}{|c|}{maxpool}                                                                                             
        &                                                                    \\ 
        \hline
conv3-384                                                 & 
        \begin{tabular}[c]{@{}c@{}}conv3-256\\ conv3-256\end{tabular} & 
            \begin{tabular}[c]{@{}c@{}}conv3-128\\ conv3-128\\ x4\end{tabular} 
                \\ \hline
\multicolumn{2}{|c|}{maxpool}                                                                                             
        &                                                                    \\ 
        \hline
conv3-384                                                  & 
        \begin{tabular}[c]{@{}c@{}}conv3-512\\ conv3-512\end{tabular} & 
            \begin{tabular}[c]{@{}c@{}}conv3-256\\ conv3-256\\ x6\end{tabular} 
                \\ \hline
\multicolumn{2}{|c|}{maxpool}                                                                                             
        &                                                                    \\ 
        \hline
conv3-256                                                 & 
        \begin{tabular}[c]{@{}c@{}}conv3-512\\ conv3-512\end{tabular} & 
            \begin{tabular}[c]{@{}c@{}}conv3-512\\ conv3-512\\ x3\end{tabular} 
                \\ \hline
\multicolumn{2}{|c|}{maxpool}                                                                                             
        & avgpool                                                            \\ 
        \hline
\multicolumn{2}{|c|}{FC-4096}                                                                                             
        & \multicolumn{1}{l|}{\multirow{2}{*}{}}                             \\ 
        \cline{1-2}
\begin{tabular}[c]{@{}c@{}}FC-4096\\ FC-4096\end{tabular} & 
    \multicolumn{1}{c|}{FC-4096}                                  & 
        \multicolumn{1}{l|}{}                                              \\ 
        \hline
\multicolumn{3}{|c|}{FC-200}                                                                                                                                                                   
        \\ \hline
\multicolumn{3}{|c|}{softmax}                                                                                                                                                                  
        \\ \hline
\end{tabular}
\label{table:config}
\end{table}


We use momentum SGD~\cite{momentum} to guide the training process with momentum 
set to 0.9.  The training process is regularized by weight decay and  the 
$L_{2}$ penalty multiplier is set to $5\times10^{-4}$.  We apply a dropout 
regularization value of 0.5 to fully-connected layers.
We initialize the weights using a zero-mean normal distribution with variance 
$10^{-2}$.  The biases are initialized to $0.1$ for Alexnet and $0$ for both VGG 
and Resnet networks.
For the Alexnet and VGG models we consider training batch sizes of 64, 32 and 
16.
To train the largest network we consider, Resnet, we consider batch sizes of 
128, 64 and 32.
The 16 batch size incurs in a prohibitively expensive training process for 
Resnet and, therefore, we do not use it in our experimental campaign. 
%
%

For Alexnet we set the initial learning rate to $10^{-2}$ for the 64 batch size 
and decrease it by factors of 2 and 4 for the 32 and 16 batch sizes, 
respectively.
In the case of VGG we set the initial learning rate to $10^{-2}$ for the 64, 32 
and 16 batch sizes, as in the state-of-the-art~\cite{vgg}.  In the case of 
Resnet the learning rate is $10^{-2}$ for the batch size of 32 and 0.1 for the 
rest.  For all network models we apply exponential decay to the learning rate 
throughout the whole training process in a way the learning rate decays every 30 
batches by a factor of 0.16, as previous work suggests~\cite{alexnet2}.
For Resnet we obtain the best results when adapting precision at the Resnet building 
block level~\cite{resnet} instead of doing it in a per-layer basis.

\subsection{Implementation}
Our code is written in Python on top of Google Tensorflow~\cite{tensorflow}.
Tensorflow is a data-flow numerical library where   
computations are driven by a computational graph that defines their order. 
It supports NVIDIA's NCCL library.

To enable the use of both Bitpack and Bitunpack routines, we integrate them into 
Tensorflow using its C++ API.
Tensorflow executes the two routines before sending the weights from the CPU to 
the GPU and right after receiving the weights on the GPU side, respectively.
The Bitpack routine is implemented using the OpenMP 4.0 programming model.  
There are two versions of this routine using either Intel's AVX2 or IBM's AltiVec 
instructions, as explained in Section~\ref{sec:approx}.
Bitunpack is implemented using CUDA 8.0 and CUDA 10.0 respectively on the two 
platforms~\cite{cuda}.

\subsection{Hardware Platforms}
\label{sec:platform}
We conduct our experiments on two clusters featuring the x86 and POWER 
architectures.
The x86 machine is composed of two 8-core Intel Xeon
\textregistered E5-2630 v3 (Haswell) at 2.4 GHz and a 20 MB L3 shared cache 
memory each.  It is also equipped with two Nvidia Tesla K80 accelerators, each 
of which hosts two Tesla GK210 GPUs.
It has 128 GB of main memory, distributed in 8 DIMMs of 16 GB DDR4 @ 2133 MHz.
The 16-core CPU and the four GPUs are connected via a PCIe 3.0 x8 8GT/s.
The operating system is RedHat Linux 6.7.
Overall, the peak performance of the two 8-core sockets plus the four Tesla 
GK210 GPUs is 6.44 TFlop/s.

The POWER machine is composed of two 20-core IBM POWER9 8335-GTG at 3.00 GHz.  
It contains four NVIDIA Volta V100 GPUs.  Each node has 512 GB of main memory, 
distributed in 16 DIMMS of 32 GB @ 2666 MHz.
The GPUs are connected to the CPU devices via a NVIDIA NVLink 2.0 
interconnection~\cite{nvlink}.
The operating system is RedHat Linux 7.4.
The peak performance of the two 20-core sockets plus the four V100 GPUs is 28.85 
TFlop/s.


\begin{figure*}[!bhtp]
    \hbox{
        \centerline{
            \includegraphics[scale=0.450]{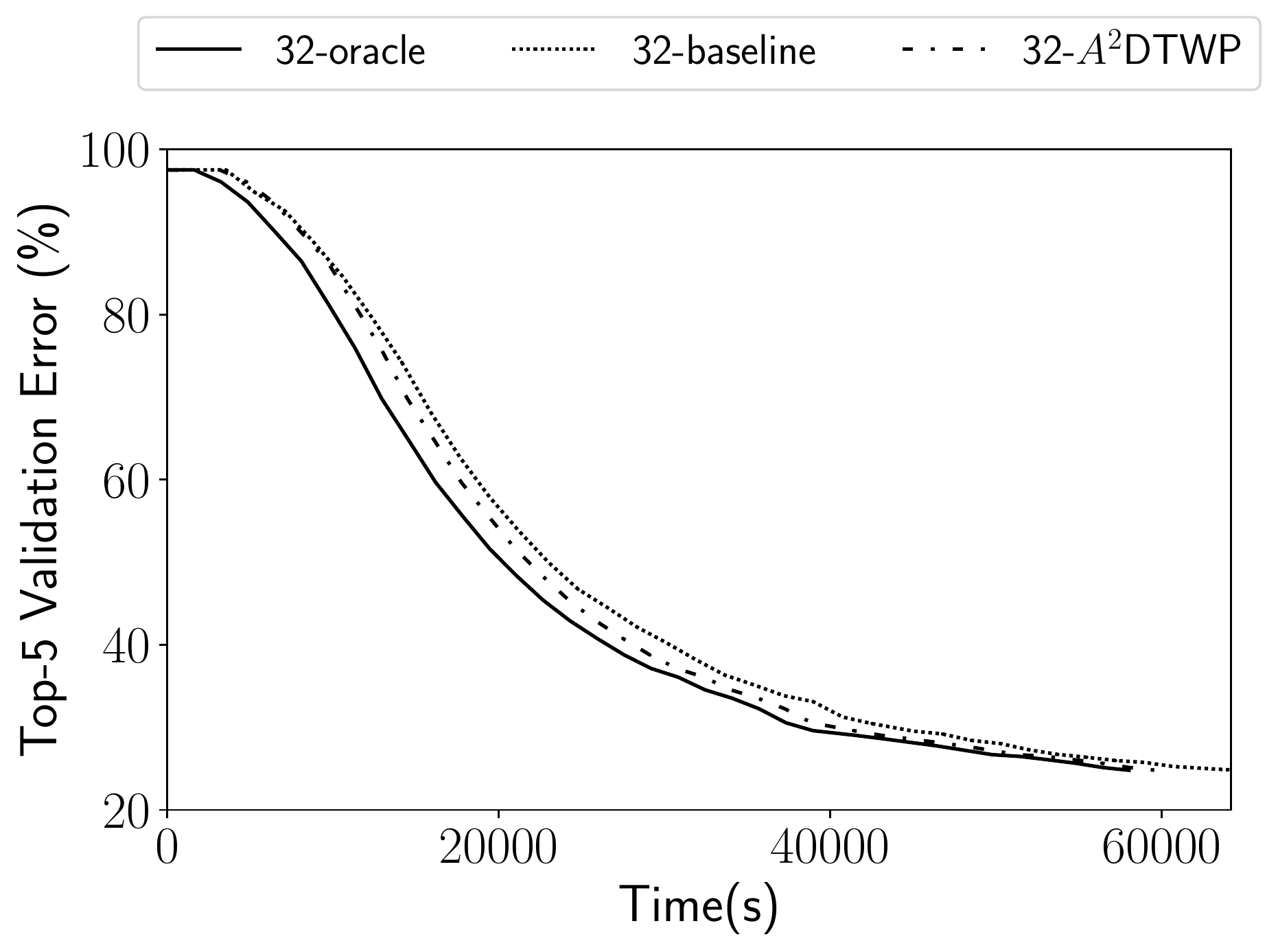}
            \includegraphics[scale=0.450]{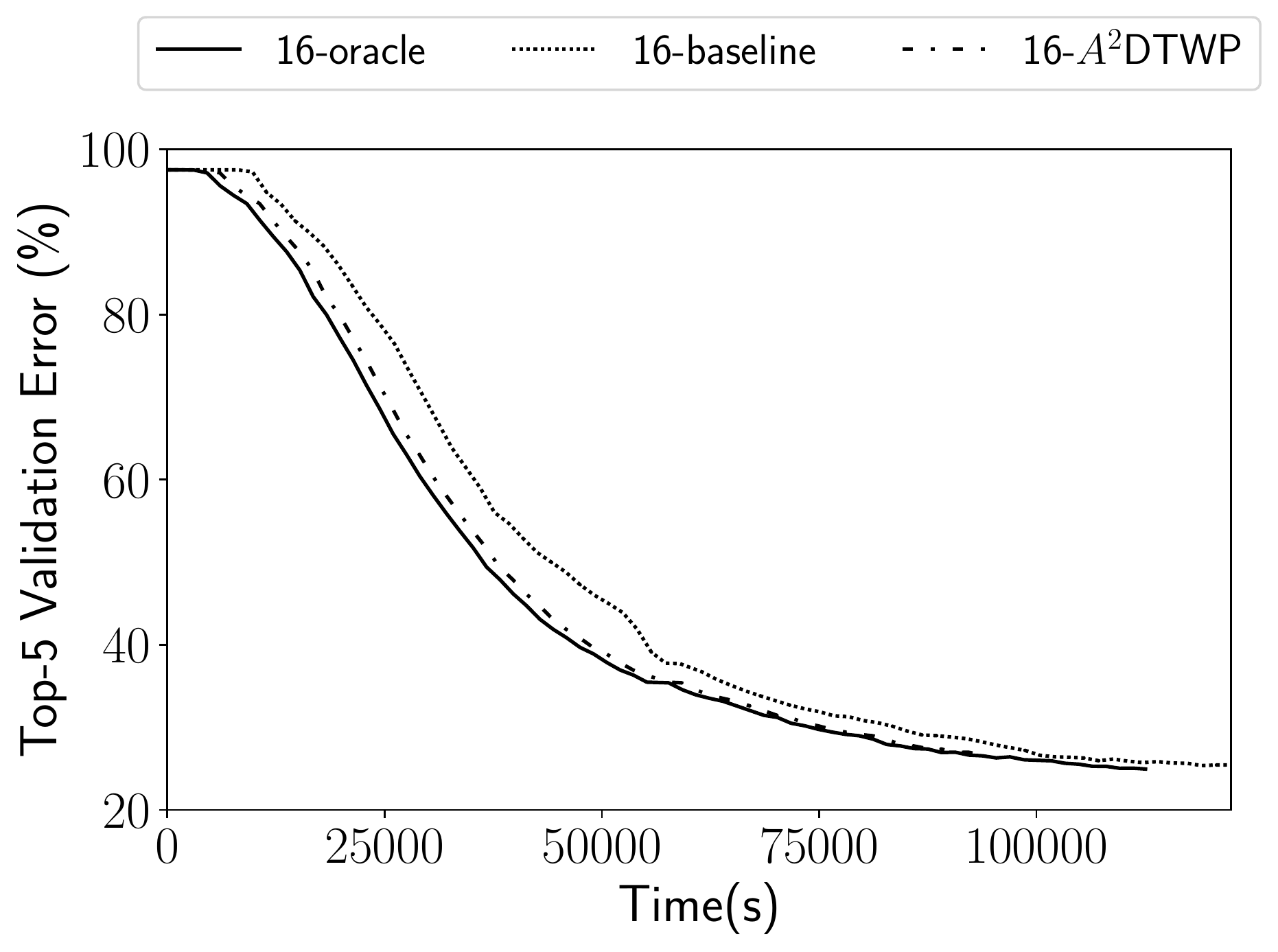}
        }
    }
    \caption{
Alexnet training considering 32 and 16 batch sizes. The two  
    plots show the top-5 validation error evolution of
    \textit{baseline}, \textit{oracle} and \textit{$A^2$DTWP}.
    \vspace{-0.5cm}
    }
        \label{alex_improv}
\end{figure*}

\section{Evaluation}
\label{sec:evalutation}
In this section we evaluate the capacity of the AWP algorithm and the ADT 
procedure to accelerate DNNs training.  We show how our proposals are able to 
accelerate the training phase of relevant DNN models without reducing the 
accuracy of the network. 

\subsection{Methodology}
\label{sec:evaluation1}

Our experimental campaign considers batch sizes of 64, 32 and 16 for the Alexnet 
and VGG models and 128, 64 and 32 for the Resnet network.
For each model and batch size, the \emph{baseline} run uses the 32-bit Floating 
Point precision for the whole training.  The data represention formats we 
consider to transfer weights from the CPU to the GPU are:
8-bit (1 bit for sign, 7 bits for exponent), 16-bit (1 bit for sign, 8 for 
exponent, 7 for mantissa), 24-bit (1 bit for sign, 8-bits for exponent and 15 
bits for mantissa) and 32-bits (1 bit for sign, 8 bits for exponent and 23 bits 
for mantissa).
We train the network models with dynamic data representation by applying the AWP 
algorithm along with the ADT procedure.
We denote this approach combining ADT and AWP as \textit{$A^2$DTWP}.  For each 
DNN and batch size, we select the data representation format that first reaches 
the 35\%, 25\% and 15\% accuracy thresholds for Resnet, Alexnet and VGG, 
respectively, and we denote this approach as \emph{oracle}.
For the case of the \emph{oracle} approach, data compression is done via ADT.
The closer \textit{$A^2$DTWP} is to \emph{oracle}, the better is the AWP 
algorithm in identifying the best data representation format.

During training we sample data in terms of elapse time and validation error 
every 4000 batches.  The total number of training batches corresponding to the 
whole ImageNet200 dataset are 16020, 8010, 4005 and 2002 for batch sizes 16, 32, 
64 and 128, respectively.
The values of $AWP$ parameters $T$, $INTERVAL$, and $N$ are determined in the 
following way:
In the case of $T$ we monitor the execution of several epochs until we observe a 
drop in the validation error.  We then measure the average change, considering 
all layers, of weights' $l^2$-norm during this short monitoring period.
The obtained values of $T$ are $-5\times10^{-2}$, $-2\times10^{-3}$ and 
$-2\times10^{-5}$ for Alexnet, VGG and Resnet, respectively.
We set the $INTERVAL$ parameter to $4000$ for both AlexNet and VGG and $2000$ 
for Resnet.  These values correspond to a single batch (for the ImageNet200 
dataset and batch sizes 64 and 128) and avoid premature precision switching due 
to numerical fluctuations.  We set $N$ to $8$ since the smallest granularity of 
our approach is 1 byte.
AWP initially applies 8-bit precision to all layers.
We use ImageNet200 in Sections~\ref{sec:alexnet}, ~\ref{sec:VGG}, 
~\ref{sec:Resnet}, ~\ref{sec:Average}, and ~\ref{sec:performance}.
Section~\ref{sec:ImageNet1000} uses ImageNet1000.

\subsection{Evaluation on Alexnet}
\label{sec:alexnet}
The evaluation considering the Alexnet model on the x86 system is shown in  
Figure~\ref{alex_improv}, which plots detailed results considering batch sizes 
of 32 and 16, and Figure~\ref{fig:all}, which shows the total execution time of 
the \textit{oracle} and \textit{$A^2$DTWP} policies normalized to the 
\textit{baseline} for the 64, 32 and 16 batch sizes on both the x86 and the 
POWER systems.
The two plots of Figure~\ref{alex_improv} depict how the validation error of 
the \textit{baseline}, \textit{oracle}, and \textit{$A^2$DTWP} policies evolves 
over time for the 32 and the 16 batch sizes until the 25\% accuracy is reached.

It can be observed in the left-hand side plot of Figure~\ref{alex_improv} 
how the \textit{oracle} and the \textit{$A^2$DTWP} approaches are 10.82\% and 
6.61\% faster than the baseline, respectively, to reach the 25\% top-5 
validation error when using a 32 batch size.
The right-hand side plot shows results considering a 16 batch size.  The 
improvements achieved by the \textit{oracle} and \textit{$A^2$DTWP} approaches 
are 11.52\% and 10.66\%, respectively.
This demonstrates the efficiency of the ADT procedure in compressing and 
decompressing the network weights without undermining the performance benefits 
obtained from sending less data
from the CPU device to the GPU.
It also demonstrates the capacity of AWP to quickly identify the best data 
representation format per layer.


Figure~\ref{fig:all} shows the normalized execution time of the \textit{oracle} 
and \textit{$A^2$DTWP} policies with respect to the 32-bit FP \textit{baseline} 
on the x86 and the POWER systems.  The top chart reports performance 
improvements of 10.75\%, 6.51\%, and 0.59\% for batch sizes 16, 32 and 64 in
the case of Alexnet runnig on the x86 system.  For the 64 batch size, the 
marginal gains of \textit{$A^2$DTWP} over the \textit{baseline} are due the poor 
performance of the 8-bits format employed by \textit{$A^2$DTWP} at the beginning 
of the training process.
This format does not contribute to reduce the validation error for the 64 batch 
case, which makes the \textit{$A^2$DTWP} policy to fall behind the 
\textit{baseline} at the very beginning of the training process.  Although 
\textit{$A^2$DTWP} eventually increases its accuracy and surpases the 
\textit{baseline}, it does not provide the same significant performance gains 
for Alexnet as the ones observed for batch sizes 16 and 32.

\textit{$A^2$DTWP} performance improvements on the POWER system in the case of 
Alexnet are 18.61\%, 14.25\% and 10.01\% with respect to the \textit{baseline} 
for batch sizes 16, 32 and 64, respectively. 
\textit{$A^2$DTWP} achieves better performance increases on the POWER system  
than x86 since its CPU to GPU bandwidth per GPUs flop/s ratio, 0.86 Bytes per Flop, is significantly slower than the x86 system ratio, 1.22 Bytes per Flop.  
This ratio expresses the maximum CPU to GPU bandwith per GPUs flop/s, which indicates the capacity to keep GPUs busy.
Since this capacity is smaller in the POWER system than in the x86 machine, our methodology achieves larger improvements when deployed on POWER.

\subsection{Evaluation on VGG}
\label{sec:VGG}


Figure~\ref{fig:all} illustrates the normalized execution time of 
\textit{$A^2$DTWP} and \textit{oracle} with respect to the \textit{baseline} for 
VGG considering batch sizes of 16, 32 and 64 on the x86 and POWER systems.
When applied to the VGG model on the x86 system, \textit{$A^2$DTWP} outperforms 
the 32-bit Floating Point \textit{baseline} by 12.88\%, 5.02\% and 7.31\% for 
batch sizes 64, 32 and 16, respectively.
Despite the relatively low performance improvement achieved by the \textit{$A^2$DTWP} 
technique when applied to the 32 batch size, \textit{$A^2$DTWP}
reaches substantial 
enhancements over the baseline in all considered scenarios. 

The performance improvements observed when training VGG on the POWER system are 
even higher.
\textit{$A^2$DTWP} outperforms the \textit{baseline} by 28.21\%, 20.19\% and 
11.13\% when using the 16, 32 and 64 batch sizes, respectively.
The performance improvement achieved on the POWER system are larger than the 
ones observed for x86 since it has less CPU to GPU bandwitdh per flop/s.
We observe the same behavior for Alexnet, as Section~\ref{sec:alexnet} 
indicates.

\subsection{Evaluation on Resnet}
\label{sec:Resnet}
\begin{figure}
    \centerline{\includegraphics[scale=0.65]{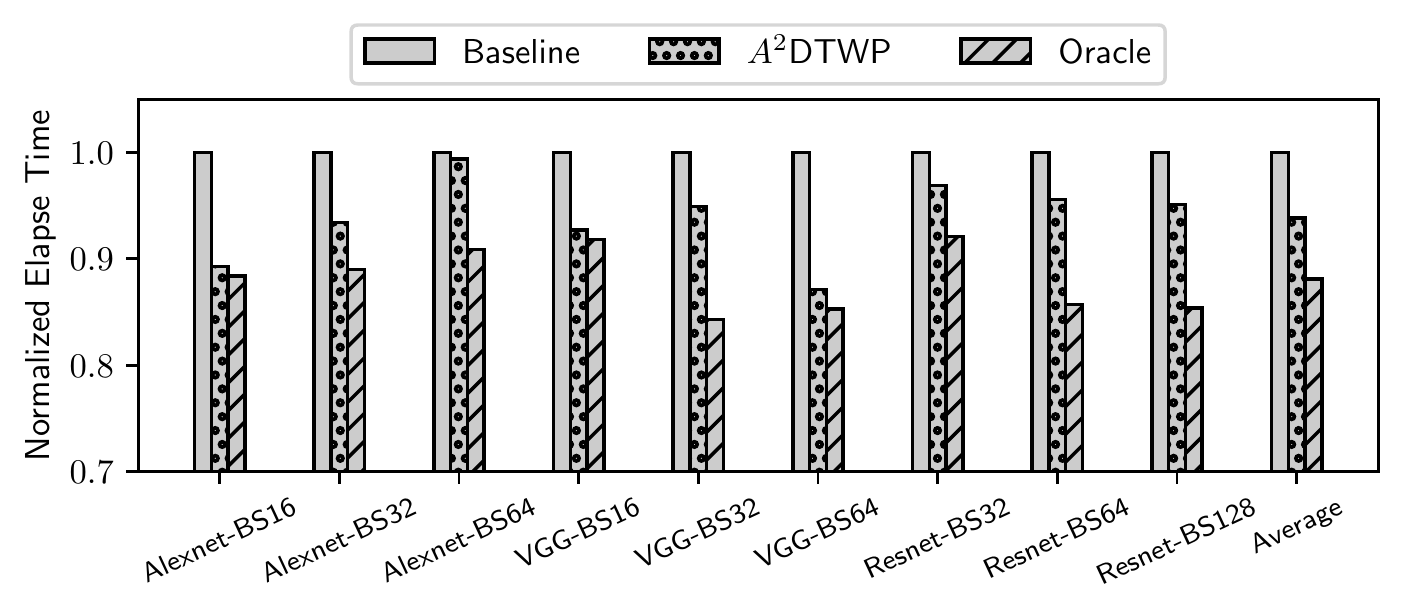}}
    \vspace{-0.2cm}
    \centerline{\includegraphics[scale=0.65]{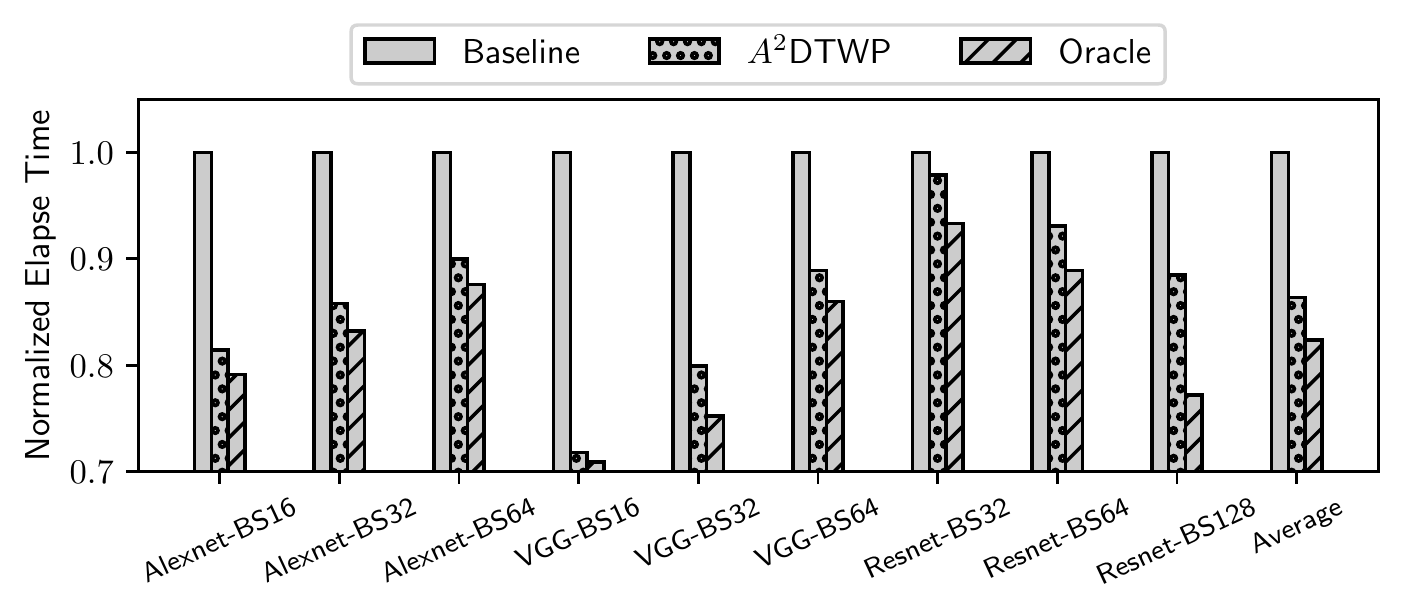}}
    \vspace{-0.2cm}
    \caption{Normalized execution times of the \textit{$A^2$DTWP} and the
    \textit{oracle} policies with respect to the baseline.  Results obtained on
    the x86 system appear in the upper plot while the evaluation on the POWER
    system appears at the bottom.}
    \label{fig:all}
    \vspace{-0.5cm}
\end{figure}

We display the normalized execution time of the \textit{$A^2$DTWP} and the 
\textit{oracle} policies when applied to the Resnet model using batch sizes of 
128, 64 and 32 in Figure~\ref{fig:all}.
On the x86 system, \textit{$A^2$DTWP} beats the 32-bit Floating Point baseline 
by 4.94\%, 4.39\% and 3.11\% for batch sizes of 128, 64 and 32, respectively, 
once a top-5 validation error of 30\% is reached.
The relatively low performance improvement achieved in the case of 32 batch size 
is due to a late identification of a competitive numerical precision, as it 
happens in the case of VGG and batch size 32.

The performance gains on the POWER system display a similar trend as the ones 
achieved on x86.  While they show the same low improvement for the 32 batch 
size, 2.12\%, \textit{$A^2$DTWP} achieves 6.92\% and 11.54\% performance gains 
for batch sizes 64 and 128, respectively.
\textit{$A^2$DTWP} achieves the largest performance improvement with respect to 
the 32-bit \textit{baseline} when run on the POWER system due to the reasons 
described in Sections~\ref{sec:alexnet} and~\ref{sec:VGG}.

\subsection{Average Performance Improvement}
\label{sec:Average}
The average performance improvement of \textit{$A^2$DTWP} over the 
\textit{baseline} considering the Alexnet, VGG and Resnet models reach 6.18\% 
and 11.91\% on the x86 and the POWER systems, respectively. 
As we explain in 
previous sections, \textit{$A^2$DTWP} obtains larger improvements on the POWER 
system than on x86 due to its smaller CPU to GPU Byte per Flop ratio.
This ratio is expected to decrease in future systems since flop/s will increase more than bandwdith, which indicates the potential of \textit{$A^2$DTWP} to achieve even larger performance gains in future systems. 

The combination of the AWP algorithm and the ADT procedure properly adapts the 
precision of each network layer and compresses the corresponding weigths with a 
minimal overhead.
The large performance improvement obtained while training deep networks on two 
high-end computing systems demonstrate the effectiveness of \textit{$A^2$DTWP}.

\subsection{Experiments with ImageNet1000}
\label{sec:ImageNet1000}
We run experiments considering ImageNet1000 to confirm they display the same
trends as executions with ImageNet200.
The experimental setup of the evaluation considering ImageNet1000 is the same
as the one we use for ImageNet200, including training and $AWT$ parameters, which are described in Sections~\ref{sec:trainingparameters} and~\ref{sec:evaluation1}.
We consider batch sizes that produce the fastest 32-bit FP training for each
one of the network models: 64, 64, and 128 for Alexnet, VGG and Resnet,
respectively.

\begin{figure}
    \centerline{\includegraphics[scale=0.65]{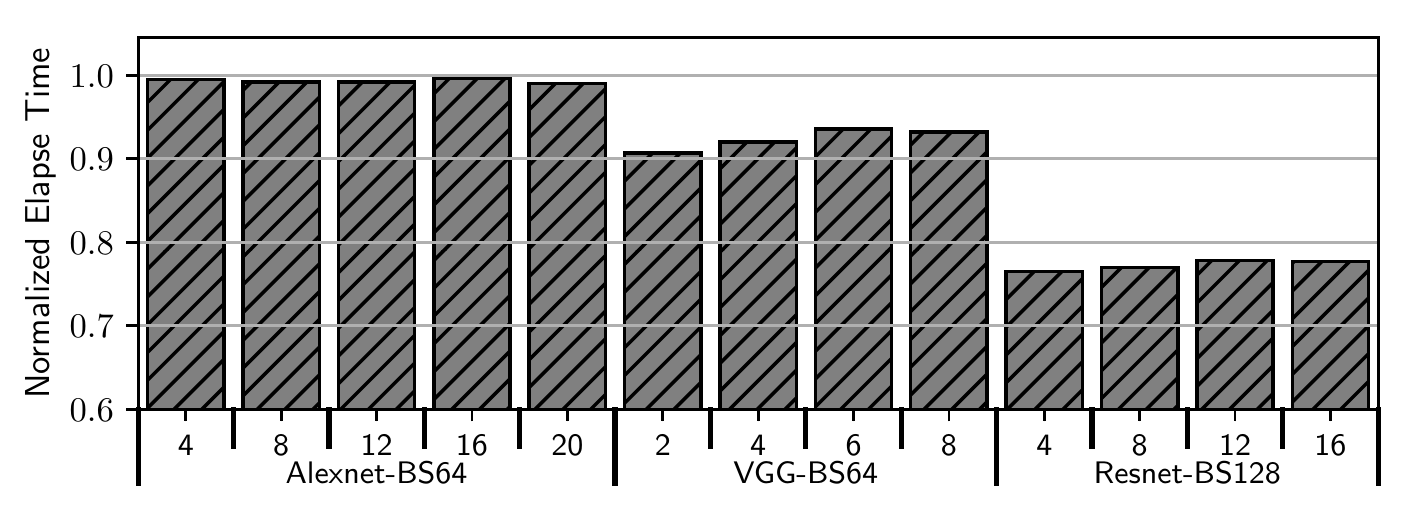}}
    \caption{Normalized execution time of \textit{$A^2$DTWP} with respect to
    \textit{baseline} considering the Imagenet1000 data set. Training for
    Alexnet, VGG and Resnet considers up to 20, 8, and 16 epochs,
    respectively.}
    \vspace{-0.5cm}
    \label{fig:ImageNet1000}
\end{figure}

Figure~\ref{fig:ImageNet1000} displays results corresponding to the
experimental campaign with ImageNet1000 on the x86 system.
In the x-axis we display different epoch counts for each one of the three
models: 4, 8, 12, 16, and 20 epochs for Alexnet; 2, 4, 6, and 8 for VGG; and 4,
8, 12, and 16 epochs for Resnet.
The y-axis displays the normalized elapsed time of \textit{$A^2$DTWP} with
respect to the the 32-bit Floating Point \textit{baseline} per each model and
epoch count.
For the case of Alexnet with batch size 64, \textit{$A^2$DTWP} is slightly
faster than the \textit{baseline} as it displays a normalized execution time of
0.995, 0.992, 0.992, 0.996, and 0.990 after 4, 8, 12, 16 and 20 epochs,
respectively.
Figure~\ref{fig:all} also reports small gains for the case of Alexnet with
batch size 64, which confirms that experiments with ImageNet1000 show very
similar trends as the evaluation with ImageNet200.
When applying \textit{$A^2$DTWP} to VGG with 64 batch size, it displays a
normalized execution time of 0.907, 0.920, 0.936, and 0.932 with respect to the
\textit{baseline} after running 2, 4, 6 and 8 training epochs, respectively.
For the Resnet example, we observe normalized execution times of 0.765, 0.770,
0.778, and 0.777 for \textit{$A^2$DTWP} after 4, 8, 12, and 16 training epochs,
respectively,
which constitutes a significant performance improvement.

In terms of validation error, both \textit{$A^2$DTWP} and \textit{baseline}
display very similar top-5 values at the end of each epoch.
For example, for the case of VGG, the Floating Point 32-bit \textit{baseline}
approach displays a validation error of 88.04\% after 2 training epochs while
\textit{$A^2$DTWP} achieves a validation error of 89.97\% for the same epoch
count, that is, an absolute difference of 1.93\%.
After 8 training epochs the absolute distance of top-5 validation errors
between \textit{$A^2$DTWP} and \textit{baseline} 
is 0.71\%.
Top-5 validation error keeps decreasing in an analogous way for both
\textit{baseline} and \textit{$A^2$DTWP} as training goes over more epochs,
although \textit{$A^2$DTWP} is significantly faster.
Our evaluation indicates that \textit{$A^2$DTWP} can effectively accelerate
training while achieving the same validation error as the 32-bit FP
\textit{baseline} when considering ImageNet1000.

\subsection{$A^2DTWP$ Performance Profile}
\label{sec:performance}
This section provides a detailed performance profile describing the effects of 
applying $A^2DTWP$ when training the VGG network model with batch size 64 on the 
x86 and POWER systems described in section~\ref{sec:platform}.
To highlight these effects we also show a performance profile of applying 32-bit 
Floating Point format during training.
The main kernels involved in the training process and their corresponding 
average execution time in milliseconds are shown in Tables
~\ref{table:performance} and~\ref{table:performance_p9}.
Each kernel can be invoked multiple times by different network layers and it can 
be overlapped with other operations while processing a batch.
Tables~\ref{table:performance} and~\ref{table:performance_p9} display for all 
kernels the average execution time of their occurrences within a batch when run 
on the x86 and the POWER systems, respectively.

Results appearing in Table~\ref{table:performance} show how time spent 
transferring data from the CPU to the GPU accelerators when applying $A^2DTWP$ 
on the x86 system, 52.27 ms, is significantly smaller than the cost of 
performing the same operation when using the 32-bit configuration, 153.93 ms.  
This constitutes a 2.94x execution time reduction that compensates the cost of 
the operations involved in the ADT routine, Bitpack and
Bitunpack, and in the AWP algorithm, the $l^2$-norm computation.
On POWER we observe a similar reduction of 3.20x in the time spent transferring
data from the CPU to the GPUs when applying $A^2DTWP$.
These reductions in terms of CPU to GPU data transfer time are due to a close to 
3x reduction in terms of weights size enabled by $A^2DTWP$.
The average execution time of operations where the $A^2DTWP$ technique plays no 
role remains very similar for the 32-bit Floating Point baseline and $A^2DTWP$ 
in both systems, as expected.
Tables~\ref{table:performance} and~\ref{table:performance_p9} indicate that 
performance gains achieved by $A^2DTWP$ are due to data motion reductions, which 
validates the usefulness of $A^2DTWP$.

Tables~\ref{table:performance} and~\ref{table:performance_p9} also display the 
overhead associated with AWP and ADT in terms of milliseconds.
The AWP algorithm spends most of its runtime computing the $l^2$-norm of the 
weights, which takes a total of 3.88 ms within a batch on the x86 system.  On 
POWER, the cost of computing the $l^2$-norm of the weights is 0.93 ms.
The other operations carried out by AWP have a negligible overhead.
The two fundamental procedures of ADT are the Bitpack and Bitunpack routines, 
which take 19.71 and 4.51 ms to run within a single batch on the x86 system.
For the case of POWER, Bitpack and Bitunpack take 10.51 and 1.11 ms, 
respectively.
Overall, measurements displayed at Table~\ref{table:performance} indicate that 
AWP and ADT constitute 1.05\% and 6.60\% of the total batch execution time, 
respectively, on x86.
On the POWER system, AWP and ADT constitute 0.54\% and 6.82\% of the total batch 
execution time according to Table~\ref{table:performance_p9}.  Figures 
~\ref{alex_improv} and ~\ref{fig:all} account for this 
overhead in the results they display.

\begin{table}
    \caption{Performance profiles of both the $A^2DTWP$ and the 32-bit Floating 
    Point approaches expressed in milliseconds on the x86 system.  We consider 
    the VGG network model with batch size 64.} \vspace{-0.25cm}
    \centering
    \begin{tabular}{|P{4.2cm}|P{1.5cm}|P{1.7cm}|}
    \hline
    & \textbf{32-bit FP} & $\mathbf{A^2DTWP}$\\
    \hline
    Data Transfer CPU$\rightarrow$GPU& 153.93 & 52.27 \\
    \hline
    Data Transfer GPU$\rightarrow$CPU& 68.51 & 73.55 \\
    \hline
    Convolution & 128.72 & 126.13\\
    \hline
    Fully-connected & 33.51 & 34.17 \\
    \hline
    Gradient update & 54.39 & 52.86\\
    \hline
    AWP ($l^2$-norm) & N/A & 3.88 \\
    \hline
    ADT (Bitpack) & N/A & 19.71 \\
    \hline
    ADT (Bitunpack) & N/A & 4.51 \\
    \hline
    \end{tabular}
    \label{table:performance}
\end{table}

\begin{table}
    \caption{Performance profiles of both the $A^2DTWP$ and the 32-bit Floating 
    Point approaches expressed in milliseconds on the POWER system.  We consider 
    the VGG network model with batch size 64.} \vspace{-0.25cm}
    \centering
    \begin{tabular}{|P{4.2cm}|P{1.5cm}|P{1.7cm}|}
    \hline
    & \textbf{32-bit FP} & $\mathbf{A^2DTWP}$ \\
    \hline
    Data Transfer CPU$\rightarrow$GPU& 39.12 & 12.21 \\
    \hline
    Data Transfer GPU$\rightarrow$CPU& 17.34 & 17.87 \\
    \hline
    Convolution & 69.78 & 71.21\\
    \hline
    Fully-connected & 12.66 & 13.51 \\
    \hline
    Gradient update & 41.29 & 42.98\\
    \hline
    AWP ($l^2$-norm) & N/A & 0.93 \\
    \hline
    ADT (Bitpack) & N/A & 10.51 \\
    \hline
    ADT (Bitunpack) & N/A & 1.11 \\
    \hline
    \end{tabular}
    \label{table:performance_p9}
\end{table}

\section{Related work}
\label{sec:Relatedworks}
A rich body of literature exists in describing the impact of using  data 
representation formats smaller than the 32-bit Floating Point standard while 
training neural networks. Previous work provides theoretical analysis on the  
learning capability under limited-precision scenarios of simple 
networks~\cite{holi93}. In recent years, researchers have shown that 
fixed-precision arithmetic is well suited for deep neural networks 
training~\cite{courbariaux14}, particularly when combined with
stochastic rounding~\cite{gupta15}. New data representation formats targeting 
dynamic and low accuracy opportunities for deep learning have been 
proposed~\cite{flexpoint17, bfloat16}.
While these approaches have a very large potential for reducing DNNs training 
costs, they do not target the data movement problem and, as such, they are 
orthogonal to the approach presented by this paper.

There is a methodology for training deep neural models using 16-bit FP numbers 
without modifying hyperparameters or losing network 
accuracy~\cite{Micikevicius2018}.
This previous approach avoids losing accuracy by keeping a 32-bit copy of 
weights, scaling the loss function to preserve small gradient updates, and
using 16-bit arithmetic that accumulates into single-precision registers.
While it also exploits the tolerance of DNN to data representation formats with 
less precision than the 32-bit FP standard, our goal is fundamentally different 
since we reduce data motion in the context of heterogeneous high-end 
architectures while this previous approach aims at reducing the computing and 
storage costs of DNN training.
This approach can be combined with \textit{$A^2$DTWP} by decompressing network 
weights to half-precision to reduce GPU computing time.
This reduction would increase the impact of data motion in the overall 
performance, which imples that the benefits of \textit{$A^2$DTWP} could be evern 
larger. 

Other approaches  
exploit model parallelism instead of data-level parallelism to orchestrate 
parallel executions of deep learning workloads~\cite{coates13, quoc11}.  
If the different 
parallel instances of this model-level parallel scheme had different precision 
requirements, \textit{$A^2$DTWP} would obtain very significant performance improvements.

Some previous approaches reduce DNNs storage and energy requirements to run 
inference on mobile devices~\cite{Han15} and achieve large graident compression 
ratios in the context of mobile device distributed trainign~\cite{Lin18}. 
While these approaches achieve very large storage reductions and substantial 
speedups, they target mobile computing.

Asynchronous SGD~\cite{jeff12} and its
variants~\cite{hogwild, zhang14} target the synchronization cost of SGD gradient updates.
Other approaches either quantize gradients to ternary levels \{-1, 0, 1\} to
reduce the overhead of gradient synchronization~\cite{sgd0}, or propose a family 
of algorithms allowing for lossy compression of gradients called Quantized SGD 
(QSGD)~\cite{sgd1}.
Techniques based on sparsifying gradient updates by removing the smallest 
gradients by absolute value~\cite{sgd2} can also reduce SGD synchronization 
costs. 
While some of these approaches apply techniques based on small data 
representation formats to reduce the synchronization costs of SGD gradient 
updates, \textit{$A^2$DTWP} targets the cost of sending DNNs weights to the GPU 
accelerators.
Therefore, these approaches are orthogonal to \textit{$A^2$DTWP} and can be 
combined with it to reduce as much as possible training communication cost.  In 
particular, techniques targeting synchronization overhead of SGD gradient updates 
can be used to reduce GPU to CPU data transfer overhead while \textit{$A^2$DTWP} 
targets CPU to GPU communication cost.

To the best of our knowledge, this paper is the first in accelerating the 
training of deep neural networks in multi-GPU high-end systems by reducing data 
motion.


\section{Conclusion}
\label{sec:conclusion}

This paper proposes $A^2DTWP$, which reduces data movement 
across heterogeneous environments composed of several GPUs and multicore CPU devices 
in the context of deep learning workloads.
The $A^2DTWP$ framework is composed of the AWP algorithm and the ADT procedure.
AWP is able to dynamically define the weights data representation format
during training. 
This paper demonstrates that AWP is
effective without any deterioration on the learning capacity of
the neural network.
To transform AWP decisions into real performance gains, 
we introduce the ADT procedure, which efficiently compresses network's weights before sending them to the GPUs. 
This procedure exploits both thread- and SIMD-level parallelism. 
By combining AWP with ADT we are able to achieve a significant performance gain 
when training network models such as Alexnet, VGG or Resnet.
Our experimental campaign considers different batch sizes and two different multi-GPU high-end systems.

This paper is the first in proposing a solution that relies on  
reduced numeric data formats to mitigate the cost of sending DNNs weights to 
different hardware devices during training.
While our evaluation targets heterogeneous high-end systems composed of several GPUs 
and CPU multicore devices, techniques presented by this paper are easily generalizable 
to any context involving several hardware accelerators exchanging large 
amounts of data.
Taking into account the prevalence of deep learning-specific accelerators in large
production systems~\cite{Jouppi2017}, the contributions of this paper are 
applicable to a wide range of scenarios involving different kinds of accelerators.

\section*{Acknowledgments}
The project OPRECOMP (website: \url{oprecomp.eu}) acknowledges the financial 
support of the Future and Emerging Technologies (FET) programme within the 
European Union's Horizon 2020 research and innovation programme, under grant 
agreement No 732631. The authors wish to thanks Dr.\ Costas Bekas --- IBM 
Research, for the outstanding support to this work. IBM, and ibm.com are 
trademarks or registered trademarks of International Business Machines 
Corporation in the United States, other countries, or both. Intel is a trademark 
or registered trademarks of Intel Corporation or its subsidiaries in the United 
States and other countries. Other product and service names might be trademarks 
of IBM or other companies.

\bibliographystyle{IEEEtran}
\bibliography{references}

\end{document}